\documentclass[12pt]{article}

\usepackage{textcomp}
\usepackage{chetnew}

\newcommand{\CH}{\mathcal{H}}
\newcommand{\CQ}{\mathcal{Q}}
\newcommand{\CB}{\mathcal{B}}
\newcommand{\CC}{\mathcal{C}}
\newcommand{\CO}{\mathcal{O}}
\newcommand{\CT}{\mathcal{T}}
\newcommand{\CI}{\mathcal{I}}
\newcommand{\CN}{\mathcal{N}}
\newcommand{\CS}{\mathcal{S}}

\preprint{RU-NHETC-2015-02}

\title{Argyres-Douglas Theories, $S^1$~Reductions, \\[1mm] and  Topological Symmetries}

\author{Matthew Buican$^{\diamondsuit}$ and Takahiro Nishinaka$^{\clubsuit}$
}

\affiliation{NHETC and Department of Physics and Astronomy \\ Rutgers University, Piscataway, NJ 08854, USA\emails{$^{\diamondsuit}$buican@physics.rutgers.edu, $^{\clubsuit}$nishinaka@physics.rutgers.edu}}

\abstract{In a recent paper, we proposed closed-form expressions for the superconformal indices of the $(A_1, A_{2n-3})$ and $(A_1, D_{2n})$ Argyres-Douglas (AD) superconformal field theories (SCFTs) in the Schur limit. Following up on our results, we turn our attention to the small $S^1$ regime of these indices. As expected on general grounds, our study reproduces the $S^3$ partition functions of the resulting dimensionally reduced theories. However, we show that in all cases---with the exception of the reduction of the $(A_1, D_4)$ SCFT---certain imaginary partners of real mass terms are turned on in the corresponding mirror theories. We interpret these deformations as $R$ symmetry mixing with the topological symmetries of the direct $S^1$ reductions. Moreover, we argue that these shifts occur in any of our theories whose  four-dimensional $\mathcal{N}=2$ superconformal $U(1)_R$ symmetry does not obey an $SU(2)$ quantization condition. We then use our $R$ symmetry map to find the four-dimensional ancestors of certain three-dimensional operators. Somewhat surprisingly, this picture turns out to imply that the scaling dimensions of many of the chiral operators of the four-dimensional theory are encoded in accidental symmetries of the three-dimensional theory. We also comment on the implications of our work on the space of general $\CN=2$ SCFTs.}

\date{May 2015}

\begin{document}
\maketitle
\toc
\newsec{Introduction}
Argyres-Douglas (AD) superconformal field theories (SCFTs) were first discovered twenty years ago at special points on the Coulomb branch of four-dimensional $\CN=2$ gauge theories where mutually non-local BPS states become massless \rcite{Argyres:1995jj, Argyres:1995xn}. These non-Lagrangian SCFTs and their generalizations \rcite{Eguchi:1996vu, Gaiotto:2010jf, Xie:2012hs} exhibit many properties that are unfamiliar from the perspective of weakly coupled theories. For example, their $\CN=2$ chiral primaries\foot{By $\CN=2$ chiral primaries, we mean primaries annihilated by all the $\CN=2$ anti-chiral Poincar\'e supercharges. These operators are sometimes referred to as \lq\lq Coulomb branch" operators.} 
generally have non-integer (rational) conformal dimensions, the $a$ and $c$ central charges of these theories can scale linearly with the rank \rcite{Xie:2013jc}, their flavor anomalies are typically non-integer, and these SCFTs are often isolated (see, however, the recent work \rcite{Buican:2014hfa}; see also \rcite{DelZotto:2015rca} for a largely complementary discussion).

While many of the above properties can be inferred simply from the existence of a Seiberg-Witten (SW) curve and a UV Lagragian from which these theories emerge in the IR, more detailed properties of these theories have long remained hidden. For example, the superconformal indices of these theories are only now being constructed and explored.\foot{See \rcite{Buican:2014qla} for a discussion of a particularly simple limit of the superconformal index for a large class of AD theories.} The main reasons for this long delay are that the superconformal $R$ symmetries of AD theories are accidental from the perspective of UV Lagrangian descriptions and that AD theories are defined by taking various subtle scaling limits. As a result, the powerful machinery of localization is not available for computing the index and for learning more about the protected spectra of these theories.

However, by generalizing a beautiful relation between $q$-deformed two-dimensional Yang-Mills theory and the Schur limit of the superconformal index \rcite{Gadde:2011ik} and by gaining inspiration from the class $\CS$ construction of AD theories \rcite{Bonelli:2011aa, Xie:2012hs}, we recently proposed and tested a closed-form expression for the index of two infinite sets of AD theories \rcite{Buican:2015ina}: the $(A_1, A_{2n-3})$ and $(A_1, D_{2n})$ SCFTs.\foot{The label $n$ is an integer related to the rank of these theories (i.e., the complex dimension of the corresponding Coulomb branch). For the $(A_1, A_{2n-3})$ SCFT, the rank is $n-2$, while for the $(A_1, D_{2n})$ theory, the rank is $n-1$. The class $\CS$ realization of these theories in terms of the $(2,0)$ theory compactified on a sphere with an irregular singularity (and a regular singularity in the case of the $(A_1, D_{2n})$ theories) was given in \rcite{Xie:2012hs}. For a recent discussion of the embedding in a four-dimensional gauge theory, see \rcite{Argyres:2012fu}.} More precisely, we suggested a form for the Schur limit of the $(A_1, A_{2n-3})$ and $(A_1 D_{2n})$ indices.

This limit of the index counts all operators of a theory that are annihilated by the $\CQ^1_{+}$ and $\tilde \CQ_{2,\dot-}$ supercharges \rcite{Gadde:2011uv}\foot{Here $-$ and $\dot+$ are values of Lorentz indices, and $1,2$ are values of $SU(2)_R$. We follow the conventions of \rcite{Buican:2014qla}.} weighted by fermion number, a fugacity, $q$, for $E-R$ where $E$ and $R$ are the scaling dimensions and $SU(2)_R$ weights of the contributing operators, and fugacities, $a_s$, for the flavor symmetries.\foot{By flavor symmetry, we mean a continuous symmetry commuting with the $\CN=2$ superconformal algebra whose corresponding current does not sit in a multiplet with higher-spin symmetries.} We can express this index formally as
\eqn{
\CI={\rm Tr}(-1)^Fq^{E-R}\prod_{s=1}^na_s^{f_s}~.
}[SchurIndex]

One of the main results of \rcite{Buican:2015ina} was to provide strong evidence that for the $(A_1, A_{2n-3})$ theory
\eqn{
\CI_{(A_1, A_{2n-3})}(q;a)=\CN(q)\sum_R[{\rm dim} \ R]_q\tilde f_{R}^{(n)}(q; a)~,
}[A1A2n3Schur]
where $\CN(q)\equiv\prod_{k=2}^{\infty}(1-q^n)^{-1}$, $[k]_q\equiv(q^{k\over2}-q^{-{k\over2}})/(q^{1\over2}-q^{-{1\over2}})$, and $R$ runs over all the irreducible representations of $su(2)$ (with dimension ${\rm dim} \ R$). The factor $\tilde f_R^{(n)}(q;a)$ (interpreted as a wave function for the irregular singularity in the class $\CS$ construction) is given by
\eqn{
\tilde f_R^{(n)}(q;a)\equiv{q^{nC_2(R)}\over(q;q)_{\infty}}{\rm Tr}_R\ \left[a^{2J_3}q^{-n(J_3)^2}\right]~,
}[irregsingwvfn]
where $(q;q)_{\infty}=\prod_{n=1}^{\infty}(1-q^n)$. In \eqref{irregsingwvfn}, $J_3$ and $C_2(R)=({\rm dim}R)({\rm dim}R-1)/4$ are the Cartan generator and quadratic Casimir of the $su(2)$ representation.

Another main result of \rcite{Buican:2015ina} was to provide strong evidence that for the $(A_1, D_{2n})$ theory
\eqn{
\CI_{(A_1, D_{2n})}(q;a, b)=\sum_R\tilde f_{R}^{(n)}(q; a)f_R(q;b)~.
}[A1D2nSchur]
Here, $\tilde f$ is as in \eqref{irregsingwvfn}, while the remaining factor (interpreted as the wave function of the regular singularity) is defined as
\eqn{
f_R(q;b)\equiv P.E.\left[{q\over1-q}\chi_{\rm adj}^{su(2)}(b)\right]\chi_R^{su(2)}(b)~,
}[regsingwvfn]
where $\chi_R^{su(2)}={\rm Tr}_R\left[x^{2J_3}\right]$ is the character for the representation $R$, and the \lq\lq plethystic exponential" is defined as $P.E.\left[F(q;a_1,\cdots, a_{\ell})\right]\equiv\exp\left(\sum_{k=1}^{\infty}{F(q^k;a_1^k,\cdots, a_{\ell}^k)\over k}\right)$.

In this paper, we will concern ourselves with studying the three-dimensional limit of \eqref{A1A2n3Schur} and \eqref{A1D2nSchur}. On general grounds, this limit must reproduce the $S^3$ partition functions of the corresponding theories reduced on a circle \cite{Dolan:2011rp, Gadde:2011ia, Imamura:2011uw}. However, this limit is subtle, and we will drop certain divergent pre-factors that were described in \rcite{Buican:2015ina}. While these pre-factors contain interesting data about the four-dimensional theory (for example, the $a-c$ conformal anomaly), they do not play a role in what follows.

As we will see in much greater detail below, the three-dimensional limits of \eqref{A1A2n3Schur} and \eqref{A1D2nSchur} contain some surprises. For one, we will find that the resulting $S^3$ partition functions are constructed with respect to $R$ symmetries that generally include mixing with the topological symmetries of the $S^1$ reductions. This mixing turns out to encode data about the scaling dimensions of the $\CN=2$ chiral primaries of the four-dimensional AD theories. This result is somewhat unexpected since the Schur limit does not receive contributions from such operators. However, the pole structure of the index turns out to know something about these operators and implies certain relations between the physics on the Coulomb branch (recall that the vevs of $\CN=2$ chiral primaries parameterize the Coulomb branch) and the physics of Schur operators. This discussion can also be taken as further evidence for the simplicity of the AD theories we consider (since the Coulomb branch spectrum is not completely independent data).

In fact, the $S^1$ reductions of our theories themselves are very simple: they can also be described by the long-distance limits of renormalization group (RG) flows from certain asymptotically free Abelian theories in three dimensions. In terms of the variables of these Abelian theories, we find another small surprise: the $\CN=2$ chiral primaries of the AD theories map to monopole operators in three dimensions.\foot{In terms of the mirror three-dimensional theories, these monopole operators are standard matter operators.} While this mapping is somewhat unusual from the perspective of the reduction of four-dimensional Lagrangian theories, it is not completely surprising in our case. Indeed, our RG flows from four dimensions never pass through the weak-coupling limit of the corresponding three-dimensional Abelian theories.\foot{We thank N.~Seiberg for a discussion of this point.} In fact, we expect generalizations of our operator map between four-dimensional chiral primaries and three-dimensional monopole operators to apply to many more general theories.

The plan of this paper is as follows. In the next section, we introduce a main tool used in our subsequent analysis and explain how it manifests itself in the study of the index. This object is an interpolating $R$ symmetry that exists as we flow from four dimensions in the UV to three dimensions in the IR when we put our AD theories on a circle. In the UV, this $R$ symmetry is the four-dimensional $U(1)_R\subset U(1)_R\times SU(2)_R$ symmetry, while in the IR it is an $R$ symmetry that can mix with the topological symmetries of the long-distance three-dimensional theories. In section 3, we argue that this mixing occurs if the four-dimensional $U(1)_R$ symmetry does not obey an $SU(2)$ quantization condition. We give some examples of theories in which we believe this mixing does not occur (including the $(A_1, D_4)$ theory). In sections 4 and 5 we then apply our formalism to the $(A_1, A_{2n-3})$ SCFTs and the remaining $(A_1, D_{2n})$ theories. We describe the resulting operator maps involving the $\CN=2$ chiral primaries in four dimensions and the resulting monpole operators in three dimensions. In section 6 we comment on potential completions of this operator map, and in section 7 we conclude and mention several open problems.

\newsec{The interpolating $R$ symmetry}
Our theories of interest are four-dimensional $\CN=2$ SCFTs. As such, their $R$ symmetry is $SU(2)_R\times U(1)_R$. One particularly important class of operators below is the set of $\CN=2$ chiral primaries. An operator, $\CO$, in this class satisfies
\eqn{
\left[ Q^i_{\alpha},\CO\right]=0~,
}[CoulombOp]
where $i=1,2$ is an $SU(2)_R$ index, and $\alpha=1,2$ is a left-handed Lorentz index. $\CO$ is charged under $U(1)_R$ but is a singlet under $SU(2)_R$. Moreover, the scaling dimension of $\CO$, $E(\CO)$, is determined by its $U(1)_R$ charge, $r(\CO)$, via (we are following the normalization conventions of \rcite{Buican:2014qla})
\eqn{
E(\CO)=-r(\CO)~.
}[CoulombOpr]
If the theory has a Coulomb branch, it can be parameterized by vevs of these types of operators.

Another interesting set of protected operators are scalar primaries of short mutliplets that are charged under $SU(2)_R$ but are neutral under $U(1)_R$, $\CO^{i_1\cdots i_{2k}}$ (the $i_a=1,2$ are symmetrized $SU(2)_R$ indices). These operators have dimension
\eqn{
E(\CO^{i_1\cdots i_{2k}})=j_R(\CO^{i_1\cdots i_{2k}})=R(\CO^{1\cdots 1})=k~,
}[HiggsOp]
where $j_R$ is the total $SU(2)_R$ spin, and $R$ is the $SU(2)_R$ Cartan. In our conventions, the highest-weight components of such operators satisfy
\eqn{
\left[\CQ^1_{\alpha}, \CO^{1\cdots1}\right]=\left[\tilde\CQ_{2\dot\alpha}, \CO^{1\cdots1}\right]=0~.
}[HHiggs]
Note that the lowest-weight components, $\CO^{2\cdots2}$, satisfy
\eqn{
\left[\tilde\CQ_{1\dot\alpha}, \CO^{2\cdots2}\right]=\left[\CQ^2_{\alpha}, \CO^{2\cdots2}\right]=0~.
}[LHiggs]
Vevs of operators of the type given in \eqref{HHiggs} and \eqref{LHiggs} parameterize the Higgs branch (if one exists).

When we write down the superconformal index of an AD theory, we are, roughly speaking, placing the SCFT on $S^3\times S^1$ with twisted $S^1$ boundary conditions and computing the partition function \rcite{Romelsberger:2005eg,  Kinney:2005ej, Festuccia:2011ws, Aharony:2003sx} (we will only consider the case of the round $S^3$ in this paper). The resulting curvature couplings give rise to non-conformal terms. However, we preserve a $U(1)_R$ symmetry and a Cartan $I_3^R\subset SU(2)_R$ subgroup. In the flat-space limit, these symmetries become, respectively, the elements $r$ and $R$ of the four-dimensional superconformal $R$-symmetry described above.

Let us now consider the regime of small $S^1$. To that end, we fix the $S^3$ to have unit radius and define $\beta=2\pi r_1$, where $r_1$ is the radius of the $S^1$. Then we take the limit $\beta\to0$. In this regime, our theories can effectively be thought of as three-dimensional.\foot{In this limit, the index generally develops an essential singularity in a pre-factor that is governed by the linear combination of anomaly coefficients $a-c$. In the case of our AD theories, this pre-factor was studied in \rcite{Buican:2015ina} and will be mentioned in passing below. When writing the $S^3$ partition function, we strip off these pre-factors, and they do not play an important role in our discussion.}

Three-dimensional $\CN=4$ superconformal theories on ${\bf R}^3$ have an $SU(2)_L\times SU(2)_R$ superconformal $R$ symmetry. The $\CN=2\subset\CN=4$ superconformal $R$-symmetry, $r_{\CN=2}^{3d}$, is the $U(1)_R$ Cartan subgroup of the diagonal $SU(2)_D\subset SU(2)_L\times SU(2)_R$. When we consider the partition function of the theory on $S^3$, we preserve this diagonal $r_{\CN=2}^{3d}$. More generally, we can consider placing the theory on $S^3$ by coupling it to the current multiplet for any $R$ symmetry related to the superconformal one by mixing with the $U(1)$ flavor symmetries of the three-dimensional theory \rcite{Jafferis:2010un, Hama:2010av, Festuccia:2011ws}.\foot{The superconformal $R$ symmetry maximizes the free energy of the theory on $S^3$ \rcite{Jafferis:2010un}.} These mixings, which appear as imaginary partners of real mass parameters and Fayet-Iliopoulos (FI) terms in the $S^3$ partition function, $Z_{S^3}$ \rcite{Festuccia:2011ws}, will play an important role in what follows (our conventions for the real and imaginary parts of the parameters appearing in $Z_{S^3}$ are opposite those in \rcite{Festuccia:2011ws}).\foot{Recall that FI terms are the real mass parameters for topological symmetries in ${\bf R}^3$ \rcite{Aharony:1997bx}.}

Therefore, if we start from the superconformal index in four-dimensions, and we take the $\beta\to0$ limit, the most general $R$ symmetry we can expect to appear in $Z_{S^3}$ is
\eqn{
r^{3d}=r_{\CN=2}^{3d}+c^a\cdot T_a^{\CC}+h^i\cdot T_i^{\CH}~,
}[ZS3genR]
where the $c^a$ and $h^i$ are real constants, the $T_a^{\CC}$ are generators for $U(1)$ symmetries acting on operators charged under $SU(2)_L$, and the $T_i^{\CH}$ are generators for $U(1)$ symmetries acting on operators charged under $SU(2)_R$. The superscripts \lq\lq$\CC$" and \lq\lq$\CH$" stand for \lq\lq Coulomb branch" and \lq\lq Higgs branch" respectively, since chiral primaries charged under these symmetries may (sometimes) acquire vevs that parameterize branches of these two types (we call these latter operators \lq\lq Higgs branch" or \lq\lq Coulomb branch" operators; note that we can in principle consider theories which also have, in the same duality frame, twisted cousins of Higgs branch and Coulomb branch operators, i.e., operators with opposite quantum numbers under $SU(2)_L\times SU(2)_R$ \rcite{Kapustin:1999ha}; however, we will not consider such theories in this paper).\foot{In four dimensions, $\CN=2$ chiral primaries (recall that the Coulomb branch is parameterized by vevs of such operators), $\CO_i$, cannot be charged under flavor symmetries. The reason is that flavor symmetry multiplets have primaries of $SU(2)_R$ spin one and are therefore forbidden from appearing in the $\CO_i\CO_{\bar i}^{\dagger}$ OPE \rcite{Buican:2013ica}. On the other hand, in three dimensions, $SU(2)_L$-charged chiral primaries (the three-dimensional analog of the four-dimensional $\CN=2$ chiral primaries) are charged under $SU(2)_L$, and current multiplets corresponding to the $\CT_a^{\CC}$ (and their non-Abelian partners) have primaries of $SU(2)_L$ spin one. Therefore, it follows that $SU(2)_L$-charged chiral primaries can be charged under the corresponding symmetries.} Note that the dimensions of such operators are given by their $R$ symmetry quantum numbers. In particular, \eqref{CoulombOpr} and \eqref{HiggsOp} are replaced by
\eqn{
E(\CO_{\CC}^{a_1,\cdots a_{2k}})=j_L(\CO_{\CC}^{a_1,\cdots a_{2k}})=k~, \ \ \ E(\CO_{\CH}^{i_1,\cdots, i_{2\ell}})=j_R(\CO_{\CH}^{i_1,\cdots, i_{2\ell}})=\ell~,
}[CH3ddims]
where $a_j$ and $i_k$ are symmetrized $SU(2)_L$ and $SU(2)_R$ indices respectively, $\CO_{\CC}^{a_1,\cdots a_{2k}}$ is an $SU(2)_L$-charged primary (a subset of these operators parameterize the Coulomb branch if it exists), and $\CO_{\CH}^{i_1,\cdots, i_{2\ell}}$ is an $SU(2)_R$-charged primary (a subset of these operators parameterize the Higgs branch if it exists).

Using the fact that the $\CT_i^{\CH}$ can only act on $SU(2)_R$-charged operators and the fact that the $SU(2)_R$ Cartan is preserved on $S^3\times S^1$, we expect the following identification of symmetries as we go from four to three dimensions via the RG flow in the $\beta\to0$ limit
\eqn{
R\to I_3^{R, 3d}+h^i\cdot T_i^{\CH}~,
}[SU2Rflow]
where $I_3^{R, 3d}$ is the Cartan of $SU(2)_R\subset SU(2)_L\times SU(2)_R$. By similar reasoning for $SU(2)_L$-charged operators and the action of $\CT_a^{\CC}$ (and using the fact that $r$ is preserved), we expect 
\eqn{
r\to I_3^{L, 3d}+c^a\cdot T_a^{\CC}~,
}[SU2Lflow]
where $I_3^{L, 3d}$ is the Cartan of $SU(2)_L\subset SU(2)_L\times SU(2)_R$.\foot{The equations \eqref{SU2Rflow} and \eqref{SU2Lflow} can be thought of as following from the map
\eqn{
r+R\to r_{\CN=2}^{3d}+c^a\cdot T_a^{\CC}+h^i\cdot T_i^{\CH}~,
}[RSymmap]
where the four-dimensional $R$ symmetry on the LHS of \eqref{RSymmap} acts non-trivially only on $Q_{1\alpha}$ and $\tilde Q^{1\dot\alpha}$.
}

In the cases of interest below, the resulting long-distance three-dimensional theories will be interacting $\CN=4$ SCFTs that can also be described as the IR limits of certain Abelian gauge theories with fundamental and bifundamental matter (recall, however, that our RG flows from the AD theories never pass through the weakly coupled limit of these gauge theories). Let $Z_{S^3}(u^a,v^i)$ be the resulting $S^3$ partition function for such a theory. The variables $u^a$ and $v^i$ are complex parameters whose real parts are the FI parameters and real mass parameters \rcite{Festuccia:2011ws} (as mentioned above, our conventions differ from those in \rcite{Festuccia:2011ws}; in particular, up to a real overall constant, we have $v^s\sim i v_{s}$, where $v_s$ is the background vector multiplet vev in \rcite{Festuccia:2011ws}). On the other hand, the imaginary parts of these variables parameterize the mixing of the manifest $U(1)_R$ symmetry on $S^3$ with the Coulomb branch and Higgs branch symmetries (the discussion in \rcite{Festuccia:2011ws} is at the level of $\CN=2$ SUSY). Note that unlike ${\rm Re}(u^a)$ and ${\rm Re}(v^i)$, the imaginary parts are not parameters of the flat-space theories. If we turn off the real masses and FI parameters (i.e., we are at the critical point of the flat-space theory), the $S^3$ partition function we should compare with the $\beta\to 0$ limit of the four-dimensional superconformal index is
\eqn{
Z_{S^3}(u^a, v^i)|_{u^a=ic^a, v^i=ih^i}~.
}[3dred]
Here the imaginary parts of $u^a$ and $v^i$ are turned on because, if we start from the four-dimensional superconformal index, the $R$-symmetry that couples to the theory on $S^3$ is \eqref{ZS3genR}. Under mirror symmetry, we have the following identification
\eqn{
Z_{S^3}(u^a, v^i)_{u^a=ic^a,v^i=ih^i}\leftrightarrow Z_{S^3}^{3dm}(\tilde u^i,\tilde v^a)_{\tilde u^i=ih^i,\tilde v^a=ic^a}~,
}[3dredmirror]
since the Higgs and Coulomb branch data are exchanged. We will find it simpler in our work below to compute the partition function in the mirror and compare this result with the $\beta\to0$ limit of our superconformal index. In particular, the signature of mixing with Coulomb branch symmetries in the direct $S^1$ reduction will be mixing with Higgs branch symmetries in the mirror.

\newsec{The $SU(2)$ quantization condition}
In \eqref{SU2Lflow}, we saw that when flowing from four dimensions to three dimensions (as $\beta\to0$), the $U(1)_R\subset U(1)_R\times SU(2)_R$ generator, $r$, can map to a linear combination of the Cartan generator of the $SU(2)_L$ $R$-symmetry, $I_3^{L, 3d}$, and various topological symmetries that act on the three-dimensional Coulomb branch.

How do we know when there is necessarily non-trivial mixing of the $R$ symmetry with the topological symmetries? One situation in which this mixing must occur is when the four-dimensional $U(1)_R$ symmetry does not obey an $SU(2)$ quantization condition. More precisely, we claim $r$ can flow to $I_3^{L, 3d}$ only if
\eqn{
r(\CO)=n_{\CO}\cdot I_3^{L, 3d}|_{\rm min}={n_{\CO}\over2}~,
}[RQuantCond]
where $\CO$ is any four-dimensional operator that does not flow to zero in the three-dimensional SCFT, $n_{\CO}$ is an integer that depends on $\CO$, and $I^{L,3d}_3|_{\rm min}$ is the minimal absolute value of the $SU(2)_L$ weight of an operator in the IR SCFT.\foot{Note that the operators, $\CO$, in \eqref{RQuantCond} need not transform as parts of short multiplets.} Clearly $I_3^{L, 3d}|_{\rm min}={1\over2}$, since there are three-dimensional supercurrents with weight $1/2$. If the condition in \eqref{RQuantCond} is violated by some $\mathcal{O}$, then there must be mixing with some other symmetries, since, by definition, the Cartan of $SU(2)_L$ satisfies an $SU(2)$ quantization condition. These additional symmetries are topological symmetries of the three-dimensional theory.

From this discussion, we expect that generic AD theories (and generic $\CN=2$ SCFTs) will have non-trivial mixing in \eqref{SU2Lflow}, i.e. there will be some $c^a\ne0$. Indeed, these theories generically have $\CN=2$ chiral primaries of non-integer and, more importantly, non-half-integer dimension. By \CoulombOp, these non-half-integer dimension operators translate into non-half-integer $U(1)_R$ charges. Since the four-dimensional Coulomb branch is embedded non-trivially in the three-dimensional Coulomb branch, it follows that these operators cannot flow to zero in three dimensions (at least the ones whose vevs parameterize the moduli space of the flat-space theory). Therefore, we expect that we will find $u^a=ic^a\ne0$ as in \eqref{3dred}. Alternatively, in the mirror description, we will find $\tilde v^a=ic^a\ne0$ (see \eqref{3dredmirror}). We will find ample evidence for this picture below.

\subsec{Some theories obeying the $SU(2)$ quantization condition}
Before discussing our AD theories of interest, which, as we have explained, should generically violate \eqref{RQuantCond}, let us first mention some theories that apparently do satisfy this condition. Indeed, many theories considered in the literature seem to satisfy \eqref{RQuantCond} under some assumptions we will discuss. This fact might explain why $R$ symmetry mixing with topological symmetries has not (to our knowledge) been observed in the $\beta\to0$ limit of the superconformal index before.

One set of examples that satisfy \eqref{RQuantCond} are the Lagrangian SCFTs in four dimensions. For example, the $\CN=2$ chiral primaries are Casimirs of the gauge group and hence have integer $U(1)_R$ charge. Although we cannot be sure that there are not operators which violate \eqref{RQuantCond} at some point on the conformal manifold of these theories (recall that the operators subject to our quantization condition need not be protected operators), we find it unlikely.

Upon reducing to three dimensions, a Lagrangian SCFT maps to the three-dimensional gauge theory with the same gauge group and matter content (however, in three dimensions, there is no longer a marginal coupling).\foot{For a comparison of the index and the corresponding $Z_{S^3}$ in the case of $SU(2)$ gauge theory with $N_f=4$, see \rcite{Gadde:2011ia}.} The four-dimensional Casimirs of the gauge group (i.e., the $\CN=2$ chiral primaries built out of the adjoint chiral multiplets) map to Casimirs of the gauge group in three dimensions. This fact is compatible with our discussion, because the adjoint chiral multiplets in three dimensions, $\Phi_i$, also have scaling dimension one (in the Abelian case, this scaling dimension follows from the fact that the $\Phi_i$ are related by $\CN=4$ SUSY to a topological current of canonical dimension).

Let us now consider some more interesting examples of theories that apparently satisfy \eqref{RQuantCond}. After discussing these theories, we will turn our attention to theories that violate this condition.

\subsubsec{The $T_N$ theories}
In \rcite{Gaiotto:2009we}, Gaiotto constructed an important class of four-dimensional $\CN=2$ SCFTs that can be engineered by putting the six-dimensional $(2,0)$ theory on a sphere with three full punctures. The Schur limit of the superconformal index of the $T_N$ theories was constructed in \rcite{Gadde:2011ik} (these results were  generalized away from the Schur limit in \rcite{Gadde:2011uv, Gaiotto:2012xa}), and takes the form
\eqn{
\CI(q, a_1, a_2, a_3)=\CN_{\rho_1,\rho_2,\rho_3}(q)\sum_{\lambda}{1\over[{\rm dim}R_{\lambda}]_q}\prod_{s=1}^3{\rm exp}\left[\sum_{n=1}^{\infty}{q^n\over1-q^n}{1\over n}\chi_{\rm adj}(a_s^n)\right]\chi_{R_{\lambda}}(a_s)~,
}[TNIndex]
where $\lambda$ is a label for representations, $R_{\lambda}$, of $SU(N)$, $\chi_{R_{\lambda}}$ is the corresponding character, and $[{\rm dim}R_{\lambda}]_q$ is the $q$-deformed dimension
\eqn{
[{\rm dim}R_{\lambda}]_q=\prod_{i<j}{[\lambda_i-\lambda_j+j-i]_q\over[j-i]_q}~.
}[dimR]
Note that the $\lambda_i$ in \eqref{dimR} are just the lengths of the rows of the Young diagram corresponding to $R_{\lambda}$.

The authors of \rcite{Nishioka:2011dq} took the $\beta\to0$ limit of \eqref{TNIndex} (dropping divergent quantities that do not affect $Z_{S^3}$) and matched it onto the three-dimensional mirror partition function for the corresponding \lq\lq star-shaped" quiver gauge theory \rcite{Benini:2010uu} (see also the discussion in \rcite{Benvenuti:2011ga}). In particular, they found that the sum over representations in \eqref{TNIndex} is replaced by an integral over a gauge group for a diagonal flavor symmetry of three linear quiver \lq\lq tails" that comprise the IR mirror theory
\eqn{
Z_{S^3}^{\CT_N, 3dm}=\int dm\Delta^2(m)\prod_{i=1}^3Z_i(m^a)~,
}[3dTN]
where $Z_i$ is the partition function of the $i^{\rm th}$ quiver tail gauge theory, $m^a$ are the fugacities for the diagonal symmetry, $\Delta(m)=\prod_{a<b}\sinh\pi(m^a-m^b)$, and we have turned off the flat-space parameters in \eqref{3dTN} (i.e., we are in the superconformal limit of the flat-space theory).

Note that there are no $\tilde v^a$ in the mirror partition function \eqref{3dTN} and so there are no $u^a$ in the direct $S^1$ reduction. This result implies that the IR limit of the $r$ symmetry in \eqref{SU2Lflow} is just $I_3^{L,3d}$. We claim this discussion is compatible with our $SU(2)$ quantization condition in \eqref{RQuantCond}. While we have not checked this claim for all the operators, $\CO$, of the $T_N$ theory subject to \eqref{RQuantCond} (moreover, we are not aware of a method that would allow us to perform this check), we can already see it is true for a highly non-trivial set of operators: the $\CN=2$ chiral primaries. In the $T_N$ theories, these operators are integer dimensional and hence have integer $r(\CO)$ and $n_{\CO}$.

\subsubsec{The $(A_1, D_4)$ theory}
While it is generically true that our AD theories do not satisfy the condition in \eqref{RQuantCond}, there is an important exception: the $(A_1, D_4)$ SCFT. As in the case of the $T_N$ theories, we have not checked that \eqref{RQuantCond} holds for all operators that flow to non-trivial operators in the $S^1$ reduction of $(A_1, D_4)$. However, we have strong evidence that this is the case. Indeed, the $(A_1, D_4)$ theory has a unique $\CN=2$ chiral generator, $\CO_0$, of dimension $3/2$ (the absence of higher-spin cousins of this operator was discussed in \rcite{Buican:2014qla}). Therefore, all $\CN=2$ chiral operators satisfy \eqref{RQuantCond}.

While consistency of our above discussion does not directly demand that there be no mixing of $r$ with topological symmetries upon reduction to three dimensions (we have not investigated whether our quantization condition can be turned into an if and only if statement), such a situation is compatible with our quantization condition (and provides some relatively weak empirical evidence that our quantization condition might be a biconditional statement). To see that this there is indeed no such mixing, let us first write down the index for $(A_1, D_4)$. We can find a useful representation of this quantity by taking $n=2$ in \eqref{A1D2nSchur} and rewriting it as follows\foot{Based on the discussions of the chiral algebras of the $E_6$ theory and the $SU(2)$ theory with $N_f=4$ in \rcite{Beem:2013sza}, a natural guess for the Schur index of the $(A_1, D_4)$ theory is that it is given by the torus partition function of the $SU(3)$ affine Kac-Moody algebra at level $k=-{3\over2}$. Indeed, our expressions below and in \eqref{A1D2nSchur} (for $n=2$) coincide with this quantity \rcite{Buican:2015ina}.}
\begin{eqnarray}\label{A1D4index}
\CI_{(A_1, D_4)}(q; a, b)&=&{\left(\CI_{\rm vect}^{SU(2)}(b)\right)^{-{1\over2}}\over(q;q)_{\infty}}\sum_{k=0}^{\infty}\Big({1\over b-b^{-1}}\sum_{s=\pm1}\left[{q^{(2k+1)^2-2k^2}b^{2k+2}a^s\over1-q^{2(k+{1\over2})}ba^s}-{q^{(2k+1)^2-2k^2}b^{-2k-2}a^s\over1-q^{2(k+{1\over2})}b^{-1}a^s}\right]\nonumber\\ &+&q^{2k(k+1)}\chi_{2k+1}(b)\Big)~.
\end{eqnarray}
Let us take $q=e^{-\beta}$, $a=e^{-i\beta\zeta_1}$, and $b=e^{-i\beta\zeta_2}$ (where $a$ and $b$ are the flavor $SU(3)$ fugacities). Taking the $\beta\to0$ limit of this equation, we find (dropping a divergent and flavor-independent pre-factor discussed in \rcite{Buican:2015ina})
\begin{eqnarray}\label{beta0lim}
\lim_{\beta\to0}\CI_{(A_1, D_4)}&\equiv& \CI^{4d\to3d}_{(A_1, D_4)}\simeq {1\over\sinh2\pi\zeta_2}\sum_{k=0}^{\infty}\sum_{s_1,s_2=\pm1}{(-1)^{s_2-1\over2}\over2(k+{1\over2})+i s_1\zeta_1+i s_2\zeta_2}\nonumber\\ &=&-{i\pi\over2}{1\over\sinh2\pi\zeta_2}\left(\tanh{\pi(\zeta_1+\zeta_2)\over2}-\tanh{\pi(\zeta_1-\zeta_2)\over2}\right)~.
\end{eqnarray}

Let us now compare \eqref{beta0lim} to $Z_{S^3}$ for the $S^1$ reduction of $(A_1, D_4)$. The resulting theory has a simple mirror description consisting of the IR limit of a $U(1)_1\times U(1)_2$ $\CN=4$ gauge theory \rcite{Xie:2012hs} (see also the discussion in \rcite{Boalch:XXXXxx, Boalch:YYYYyy}). The matter fields are $X$ with charge $(1,1)$, $A$ with charge $(1,0)$, and $\hat A$ with charge $(0,1)$. $X$, $A$, and $\hat A$ have corresponding partners $Y$, $B$, and $\hat B$ of opposite charge. There is also a $U(1)$ flavor symmetry, $\tilde J$, under which $\tilde J(X)=\tilde J(A)=\tilde J(\hat A)=1/2$, and the partners have opposite charge. Note that this flavor symmetry translates into a $U(1)$ symmetry of the Coulomb branch of the direct $S^1$ reduction.\foot{It would be interesting to understand if this symmetry descends from some symmetry in four dimensions.} 

The $S^3$ partition function for this theory can be computed using the methods in \rcite{Kapustin:2009kz}
\eqn{
Z_{S^3}^{(A_1, D_4), 3dm}=\int d\sigma_1d\sigma_2{e^{\pi i(\xi_1\sigma_1+\xi_2\sigma_2)}\over\cosh\pi\sigma_1\cosh\pi(\sigma_1-\sigma_2)\cosh\pi\sigma_2}={1\over2\cosh\pi{\xi_1\over4}\cosh\pi{\xi_2\over4}\cosh\pi{\xi_1+\xi_2\over4}}~,
}[ZS3A1D4]
where the $\xi_i$ are the FI parameters of the three-dimensional theory. It is straightforward to check that
\eqn{
\CI^{4d\to3d}_{(A_1, D_4)}\simeq Z_{S^3}^{(A_1, D_4)}~.
}[IZS3A1D4]
When we write \lq\lq$\simeq$", we mean that the two sides of the relation agree up to an unimportant overall factor that is independent of the continuous parameters of the theories. Note that we have identified $\zeta_1={1\over4}(\xi_1-\xi_2)$ and $\zeta_2={1\over4}(\xi_1+\xi_2)$.

In particular, we see that the three-dimensional reduction of the index and the $S^1$ reduction of the partition function agree (up to an overall constant that is not directly relevant to our analysis) without the need to turn on an imaginary partner of the real mass parameter corresponding to  $\tilde J$ in the mirror theory. This result is therefore consistent with our above discussion. In particular, we have
\eqn{
r\to I_{3}^{L,3d}~.
}[rmap]
By \eqref{CH3ddims}, we see that the four-dimensional $\CN=2$ chiral generator, $\CO_0$, should map to an operator of dimension $3/2$ in three dimensions.

To understand which operator $\CO_0$ maps to in the $S^1$ reduction, let us again consider the mirror. In particular, we are interested in the operators that parameterize the Higgs branch of the mirror (in the flat-space limit) since they map to operators that parameterize the Coulomb branch of the direct $S^1$ reduction (in the flat-space limit). The mirror theory has an $\CN=4$ superpotential
\eqn{
W=\Phi_1(AB+XY)+\Phi_2(\hat A\hat B+XY)~.
}[N4WA1D4]
As a result, we have
\eqn{
AB=\hat A\hat B=-XY~.
}[RelatA1D4]
Let us define $\tilde\CO^{22}_{0}=XY$, $\tilde\CO^{222}_{-1/2}=BX\hat B$, $\tilde\CO^{222}_{1/2}=AY\hat A$, where the superscripts give the $SU(2)_R$ quantum numbers, and the subscripts give the symmetry quantum numbers under the $U(1)_{\tilde J}$ symmetry. In terms of these variables, we find
\eqn{
\tilde\CO^{222}_{-1/2} \tilde\CO^{222}_{1/2}=(\tilde\CO^{22}_{0})^3~.
}[HiggsMirrorA1D4]
In particular, we see that the Higgs branch of the mirror (in the flat-space limit) is ${\bf C^2}/{\bf Z_3}$. Under mirror symmetry
\eqn{
\tilde\CO^{222}_{-1/2}\leftrightarrow\CO^{222}_{-1/2}~, \ \ \ \tilde\CO^{222}_{1/2}\leftrightarrow\CO^{222}_{1/2}~,\ \ \ \tilde\CO^{22}_{0}\leftrightarrow\CO^{22}_{0}~,
}[MirrorSymm]
where the operators on the RHS of the first two mappings are monopole operators (of dimension $3/2$) parameterizing the ${\bf C^2}/{\bf Z_3}$ Coulomb branch of the direct $S^1$ reduction, and $\CO_0^{22}$ is the vector multiplet chiral scalar. The subscripts on the RHS of \eqref{MirrorSymm} are charges with respect to the topological symmetry, $J$, of the direct $S^1$ reduction ($\tilde J\leftrightarrow J$ under mirror symmetry). As a result, we see that the four-dimensional operator, $\CO_0$, maps to the following linear combination of monopole operators that parameterize the Coulomb branch of the $S^1$ reduction (in the flat space limit)
\eqn{
\CO_0\to c_+\CO^{222}_{1/2}+c_-\CO^{222}_{-1/2}~,
}[OpMapA1D4]
where $c_{\pm}$ are undetermined constants. Therefore we conclude that, unlike in the case of the reduction of a Lagrangian SCFT, $\CN=2$ chiral operators in the $(A_1, D_4)$ SCFT map to monopole operators in three dimensions.

\newsec{$S^1$ reduction of the $(A_1, A_{2n-3})$ theory}
In this section we turn our attention to one of the main theories we wish to study: the $(A_1, A_{2n-3})$ theory (with $n\ge3$ so that it is interacting). As we will see, unlike the examples discussed so far, the $(A_1, A_{2n-3})$ SCFT does not satisfy the $SU(2)$ quantization condition in \eqref{RQuantCond}. As a result, we will find an intricate pattern of $R$ symmetry mixing with topological symmetries of the resulting $S^1$ reduction.

Two pieces of information about this theory are relevant to our discussion below. First, the $(A_1, A_{2n-3})$ theory has a $U(1)$ flavor symmetry for $n>3$ and an $SU(2)$ flavor symmetry for $n=3$ (the free hypermultiplet case, $n=2$, also has an $SU(2)$ flavor symmetry). Second, this theory has rank $n-2$ and a spectrum of $\CN=2$ chiral generators
\eqn{
E(\CO_{\ell})=2\left(1-{1\over n}\right)-{\ell\over n}~, \ \ \ 0\le\ell\le n-3~.
}[dimClops]
We see from this expression that, with the exception of the free hypermultiplet theory, all the $(A_1, A_{2n-3})$ theories violate the $SU(2)$ quantization condition \eqref{RQuantCond} in the $\CN=2$ chiral sector.

In order to understand the quantitative details of this picture, let us first rewrite the Schur index in \eqref{A1A2n3Schur} as follows
\begin{eqnarray}\label{A1A2n3index}
\CI_{(A_1, A_{2n-3})}(q;a)&=&{1-q\over(q;q)_{\infty}^2}\sum_{k=0}^{\infty}\Big({1\over q^{1\over2}-q^{-{1\over2}}}\sum_{s=\pm1}\left[{q^{n(2k+1)^2-2nk^2+2k+2\over2}a^s\over1-q^{(k+{1\over2})n+{1\over2}}a^s}-{q^{n(2k+1)^2-2nk^2-2k-2\over2}a^s\over1-q^{(k+{1\over2})n-{1\over2}}a^s}\right]\nonumber\\ &+& q^{nk(k+1)}{q^{k+{1\over2}}-q^{-k-{1\over2}}\over q^{1\over2}-q^{-{1\over2}}}\Big)~.
\end{eqnarray}
Taking $\beta\to0$ with $q=e^{-\beta}$ and flavor fugacity $a=e^{-i\beta\zeta}$, we find (recall that we are dropping a flavor-independent divergent pre-factor)
\begin{eqnarray}\label{beta0limA1A2n3}
\lim_{\beta\to0}\CI_{(A_1, A_{2n-3})}&\equiv& \CI^{4d\to3d}_{(A_1, A_{2n-3})}\simeq\sum_{k=0}^{\infty}\sum_{s=\pm1}\left({1\over(k+{1\over2})n+{1\over2}+is\zeta}-{1\over(k+{1\over2})n-{1\over2}+is\zeta}\right)\nonumber\\ &=& {i\pi\over n}\left(\tanh{\pi(2\zeta+i)\over2n}-\tanh{\pi(2\zeta-i)\over2n}\right)~.
\end{eqnarray}

\begin{table}
\begin{center}
     \begin{tabular}{| c | c | c | c | c | c |}
\hline   & $U(1)_1$ & $U(1)_2$ & $\cdots$ & $U(1)_{n-3}$& $U(1)_{n-2}$\cr\hline\hline
        $q_1$ & $1$ & $0$& $\cdots$ & 0 & 0\cr\hline
        $q_2$ & $1$ & $1$& $\cdots$ & 0 & 0\cr\hline
        $\vdots$ & $\vdots$ & $\vdots$& $\ddots$ & $\vdots$ & \vdots\cr\hline
        $q_{n-2}$ & $0$ & $0$& $\cdots$ & 1 & 1\cr\hline
        $q_{n-1}$ & $0$ & $0$& $\cdots$ & 0 & 1\cr\hline
      \end{tabular}
\caption{The matter fields (and charges) of the three-dimensional gauge theory that flows to the $S^1$ reduction of the $(A_1,A_{2n-3})$ theory.}
\label{AAside}
\end{center}
\end{table}
The $S^1$ reduction of the $(A_1, A_{2n-3})$ theory flows to a three-dimensional $\CN=4$ SCFT that is also the IR limit of the $U(1)^{n-2}$ gauge theory with matter content summarized in Table \ref{AAside}. The theory also has hypermultiplet partners, $\tilde q^i$, with opposite quantum numbers under the $U(1)^{n-2}$ gauge symmetry.

However, in what follows, we will find it easier to work with the mirror \rcite{Xie:2012hs,Boalch:XXXXxx, Boalch:YYYYyy}. This theory is just SQED with $N_f=n-1$. We denote the fundamental flavors as $X_I$ and the anti-fundamental $\CN=4$ partners as $Y^I$. The $S^3$ partition function for the mirror is
\eqn{
Z_{S^3}^{(A_1, A_{2n-3}),3dm}(\tilde u, \tilde v^a)|_{\tilde u=\xi, \tilde v^a=0}=\int d\sigma{e^{\pi i\xi\sigma}\over\cosh^{n-1}\pi\sigma}~,
}[ZS3A1A2n3]
where the FI parameter, $\xi$, corresponds to the real mass parameter in the direct $S^1$ reduction.
Note that this theory has an $SU(n-1)$ flavor symmetry that is accidental from the point of view of the $(A_1, A_{2n-3})$ SCFT. Under this symmetry, the $X_I$ transform as fundamentals and the $Y^I$ transform as anti-fundamentals.

We should now try to match the pole structures of \eqref{beta0limA1A2n3} and \eqref{ZS3A1A2n3}. 
To perform this comparison, we must first express $\zeta$ in terms of $\xi$. These variables are clearly related since $\zeta$ is the vev of (the zero-th component of) the background gauge field for the flavor symmetry in four dimensions, and $\xi$ is dual (by mirror symmetry) to the vev of a background real scalar coupled to the corresponding flavor symmetry in three dimensions. In fact, it turns out that the precise relation is
\eqn{
\zeta={\xi\over2}~.
}[zetaxirel]
Given this dictionary, it is straightforward to check that the pole structures in \eqref{beta0limA1A2n3} and \eqref{ZS3A1A2n3} do not match.

We claim this mismatch can be explained in terms of the RG flow of the $U(1)_R$ symmetry and the resulting mixing with topological symmetries of the three-dimensional theory. Before explaining this statement for the case of general $n$, let us consider the $(A_1, A_3)$ SCFT.

The $(A_1, A_3)$ theory has a single $\CN=2$ chiral generator of dimension $4/3$ which clearly violates \eqref{RQuantCond}. This means that there should be non-trivial $R$-symmetry mixing of the form \eqref{SU2Lflow}. Indeed, in order to find $Z_{S^3}^{(A_1, A_3)}\simeq \CI^{4d\to3d}_{(A_1, A_{3})}$ (as before, when we write \lq\lq$\simeq$", we mean that the two sides of the relation agree up to an unimportant overall factor that is independent of the continuous parameters of the theories), we must turn on an imaginary partner of the real mass in the three-dimensional mirror
\begin{eqnarray}\label{ZS3A1A3}
Z_{S^3}^{(A_1, A_{3}),3dm}(\tilde u, \tilde v)|_{\tilde u=\xi, \tilde v=\pm i/3}&=&\int d\sigma{e^{\pi i\xi\sigma}\over\cosh\pi(\sigma\mp i/6)\cosh\pi(\sigma\pm i/6)}\nonumber\\ &\simeq&{i\pi\over 3}\left(\tanh{\pi(\xi+i)\over6}-\tanh{\pi(\xi-i)\over6}\right)=\CI^{4d\to3d}_{(A_1, A_{3})}~.
\end{eqnarray}
The ambiguity in the sign of the imaginary part of $\tilde v$ reflects the fact that the partition function is invariant under an $SU(2)$ Weyl reflection.\foot{The relative factor of two appearing in $\tilde v$ versus the terms $\pm i/6$ in the arguments of the trigonometric functions on the RHS of \eqref{ZS3A1A3} is due to the fact that we take the generator of the $SU(2)$ Cartan to be $H={1\over2}{\rm diag}(1,-1)$.} Turning on the imaginary partner of the real mass term breaks the accidental flavor symmetry from $SU(2)\to U(1)$ and corresponds to turning on an imaginary partner of the topological real mass term in the direct $S^1$ reduction that breaks the Coulomb branch symmetry from $SU(2)\to U(1)$.

Using \eqref{SU2Lflow} and \eqref{3dredmirror}, we read off the following flow upon compactifying on $S^1$
\eqn{
r\to I_3^{L,3d}+{1\over3}H~,
}[RsymFlowA1A3]
where $H$ is the topological symmetry, i.e., the Cartan of the Coulomb branch $SU(2)$ symmetry. In writing \eqref{RsymFlowA1A3}, we have chosen a particular sign in \eqref{ZS3A1A3}: $Z_{S^3}^{(A_1, A_{3})}(\xi, i/3)$ (we assign the contribution from the lowest $SU(2)$ weight component of the hypermultiplet doublet to the first factor in the denominator of the integrand of \eqref{ZS3A1A3}). Under a Weyl reflection, \eqref{RsymFlowA1A3} becomes $r\to I_3^L-{1\over3}H$.

Given \eqref{RsymFlowA1A3}, we can ask how the $\CN=2$ chiral generator, $\CO_0$, is mapped under the RG flow. Since this operator has $r(\CO_0)=-4/3$, we must find a chiral operator of the same charge under the RHS of \eqref{RsymFlowA1A3}. Working in the mirror theory, we should find a chiral operator of charge $-4/3$ under $I_3^{R,3dm}+{1\over3}\tilde H$ (where $\tilde H\leftrightarrow H$ and $I_3^{R,3dm}\leftrightarrow I_3^{L,3d}$ under mirror symmetry). The chiral operators satisfying this condition are $X_2Y^1$ and $(X_1Y^2)^2$. Under mirror symmetry, these operators map to the dimension one monopole operator, $\CO^{22}_{-1}$, (this is the primary moment map for the multiplet that contains a symmetry current associated with the enhancement of the topological symmetry to $SU(2)$) and the dimension two monopole operator, $(\CO^{22}_{+1})^2$, respectively, where the subscript is the $SU(2)$ Coulomb branch symmetry weight. As a result, we see that
\eqn{
\CO_0\to c_{0,-1}\CO^{22}_{-1}+c_{0,+2}(\CO^{22}_{+1})^2~,
}[OpmapA1A3]
where $c_{0,-1}$ and $c_{0,+2}$ are constants. Under a Weyl reflection we have $\CO^{22}_{-1}\leftrightarrow\CO^{22}_{+1}$, since the Weyl reflection exchanges the positive and negative roots of $SU(2)$.

While we do not know how to compute the coefficients in \eqref{OpmapA1A3}, we can argue that $c_{0,-1}\ne0$. To understand this statement, note that, since vevs of $\CO_0$ parameterize the four-dimensional Coulomb branch (in the flat space limit), we expect this operator to flow to a three-dimensional operator whose vev parameterizes a subspace of the hyperk\"ahler Coulomb branch moduli space (in the flat space limit). Indeed, the Coulomb branch moduli space of the $S^1$ reduction (in the flat space limit) is described by 
\eqn{
\CO^{22}_{-1}\CO^{22}_{+1}=(\CO_0^{22})^2~,
}[A1A3Coulomb]
and so we expect $c_{0,-1}\ne0$ (presumably $c_{0,+2}=0$ by dimensional analysis in the flat-space limit). By an appropriate rescaling we can then rewrite \eqref{OpmapA1A3} as
\eqn{
\CO_0\to \CO^{22}_{-1}+c_{0,+2}(\CO^{22}_{+1})^2~.
}[OpmapA1A3final]

In particular, we conclude that the non-trivial scaling dimension of the $\CN=2$ chiral generator in four dimensions is encoded in the quantum numbers of the monopole operator that is related by $\CN=4$ SUSY to currents associated with the $SU(2)$ accidental symmetry in three dimensions.

Let us now consider the general case. The mirror of the $S^1$ reduction of the $(A_1, A_{2n-3})$ theory was described around \eqref{ZS3A1A2n3}. It has an $SU(n-1)$ flavor symmetry (that is accidental from the point of view of the four-dimensional theory) under which the $X_I$ and $Y^I$ transform as fundamentals and anti-fundamentals respectively. Up to unimportant Weyl reflections that generalize the previous discussion for $n=3$, we find that in order for $Z_{S^3}^{(A_1, A_{2n-3})}$ and $\CI^{4d\to3d}_{(A_1, A_{2n-3})}$ in \eqref{beta0limA1A2n3} to coincide (up to an overall constant independent of the continuous parameters) we must turn on the following $SU(n-1)\to U(1)^{n-2}$ breaking imaginary partners of the real masses in the case of even $n$ (i.e., $n=2p$)\foot{We suppress the obvious arguments of $Z_{S^3}^{(A_1, A_{4p-3}), 3dm}$ for notational simplicity.}
\begin{eqnarray}\label{ZS3A1A2m3}
Z_{S^3}^{(A_1,A_{4p-3}),3dm}&=&\int d\sigma{e^{\pi i \xi\sigma}\over\cosh\pi\sigma\cdot\prod_{\alpha=1}^{p-1}\cosh\pi(\sigma-{i\over 2p}\alpha)\cosh\pi(\sigma+{i\over 2p}\alpha)}\nonumber\\ &\simeq& {i\pi\over2p}\left(\tanh{\pi(\xi+i)\over4p}-\tanh{\pi(\xi-i)\over4p}\right)=\CI^{4d\to3d}_{(A_1, A_{4p-3})}~.
\end{eqnarray}
Similarly, in the case of odd $n$ (with $n=2p+1$), we have
\begin{eqnarray}\label{ZS3A1A2m3odd}
Z_{S^3}^{(A_1,A_{4p-1}),3dm}&=&\int d\sigma{e^{\pi i \xi\sigma}\over\prod_{\alpha=1}^{p}\cosh\pi\left(\sigma-{i\over 2(2p+1)}(2\alpha-1)\right)\cosh\pi\left(\sigma+{i\over 2(2p+1)}(2\alpha-1)\right)}\nonumber\\ &\simeq& {i\pi\over2p+1}\left(\tanh{\pi(\xi+i)\over2(2p+1)}-\tanh{\pi(\xi-i)\over2(2p+1)}\right)=\CI^{4d\to3d}_{(A_1, A_{4p-1})}~.
\end{eqnarray}

The imaginary partners of the real masses in \eqref{ZS3A1A2m3} and \eqref{ZS3A1A2m3odd} correspond to the following shift of the $\CN=2\subset\CN=4$ $R$ symmetry in the mirror
\eqn{
r^{3dm}\to I_3^{L,3dm}+I_3^{R,3dm}+{1\over n\sqrt{2}}\sum_{a=1}^{n-2}(-1)^{n+a}{\sqrt{a(a+1)}}\tilde H_a~,
}[rmapA1A2m3mirror]
where the $\tilde H_a$ are the Cartans of the $SU(n-1)$ flavor symmetry (normalized such that $2\,\text{Tr}(\tilde H_a \tilde H_b) = \delta_{ab}$).
Therefore, we conclude that in the RG flow from the four-dimensional theory to the direct $S^1$ reduction, we have
\eqn{
r\to I_3^{L,3d}+{1\over n\sqrt{2}}\sum_{a=1}^{n-2}(-1)^{n+a}{\sqrt{a(a+1)}}H_a~,
}[rmapA1A2m3]
where $\tilde H_a\leftrightarrow H_a$ under mirror symmetry. In particular, the $H_a$ are generators of the $SU(n-1)$ Coulomb branch symmetry of the direct $S^1$ reduction
\begin{eqnarray}\label{Cartan}
&&H_1={1\over2}{\rm diag}(1,-1,0,\cdots, 0)~, \  H_2={1\over\sqrt{12}}{\rm diag}(1,1,-2,0,\cdots,0)~, \ \cdots, \nonumber\\ &&H_{k}={1\over\sqrt{2k(k+1)}}{\rm diag}(1,\cdots,1, -k,0,\cdots,0)~, \ \cdots, \nonumber\\ &&H_{n-2}={1\over\sqrt{2(n-2)(n-1)}}{\rm diag}(1,\cdots,1,-(n-2))~.
\end{eqnarray}
The resulting weights for the fundamental representation are
\begin{eqnarray}\label{weights}
&&\nu_1=\left({1\over2},{1\over\sqrt{12}},\cdots, {1\over\sqrt{2(n-2)(n-1)}}\right)~, \  \nu_2=\left(-{1\over2},{1\over\sqrt{12}},\cdots, {1\over\sqrt{2(n-2)(n-1)}}\right)~, \nonumber\\ \ &&\cdots, \ \nu_{k}=\left(0,\cdots,-{k-1\over\sqrt{2(k-1)k}},\cdots,{1\over\sqrt{2(n-2)(n-1)}}\right)~, \ \cdots, \nonumber\\ &&\nu_{n-1}=\left(0,\cdots,-{n-2\over\sqrt{2(n-2)(n-1)}}\right)~.
\end{eqnarray}
From these weights, we can define the $n-2$ simple roots to be $\alpha_a=\nu_a-\nu_{a+1}$.

A particularly important set of operators parameterize the $2(n-2)$ complex dimensional Coulomb branch (in the flat space limit) via the $n-2$ equations
\eqn{
\CO^{22}_{-\alpha_a}\CO^{22}_{\alpha_a}=\CO_{0,a}^{22}\CO_{0,a+1}^{22}~,
}[SimpleRooteqns]
where the $\CO^{22}_{0,a}$ are the dimension one vector multiplet chiral scalars for the $U(1)^{n-2}$ gauge symmetry (and $\CO_{0,n-1}^{22}=-\sum_{b=1}^{n-2}\CO_{0,b}^{22}$), the $\CO^{22}_{\alpha_a}$ are monopole operators of dimension one that correspond to the simple roots of the $SU(n-1)$ symmetry (they are primaries for multiplets that contain the corresponding conserved currents), and the $\CO^{22}_{-\alpha_a}$ are monopole operators corresponding to the reflected roots. We also have the following equations
\eqn{
\CO^{22}_{{\nu_{a_0}-\nu_{a_0+\ell}}}\prod_{a=a_0+1}^{a_0+\ell-1}\CO_{0,a}^{22}=\prod_{a=a_0}^{a_0+\ell-1}\CO^{22}_{\alpha_a}~,
}[genrooteqn]
where the $\CO^{22}_{\nu_{a_0}-\nu_{a_0+\ell}}$ with $\ell>1$ are monopole operators for the positive (non-simple) roots.\foot{Under mirror symmetry, $\CO_{0,a}^{22}\leftrightarrow X_aY^a$, $\CO_{\alpha_a}^{22}\leftrightarrow X_aY^{a+1}$, $\CO_{-\alpha_a}^{22}\leftrightarrow X_{a+1}Y^{a}$, and $\CO^{22}_{{\nu_{a_0}-\nu_{a_0+\ell}}}\leftrightarrow X_{a_0}Y^{a_0+\ell}$. The operator relations discussed above can then be straightforwardly derived in the mirror theory (see also the recent discussion in \rcite{Bullimore:2015lsa}).} There is an analogous set of equations for operators corresponding to the negative roots.

Using the same arguments as around \eqref{OpmapA1A3final}, we find the following operator map
\begin{eqnarray}\label{OmapA1A2m3}
&&\CO_0\to\CO^{22}_{-\alpha_{n-2}}+\cdots~, \ \CO_{1}\to c_{1,\alpha_{n-3}}\CO^{22}_{\alpha_{n-3}}+\cdots~, \cdots~, \CO_k\to c_{1,\alpha_{n-k-2}}\CO^{22}_{(-1)^{k+1}\alpha_{n-k-2}}+\cdots~,\nonumber\\&& \cdots~, \CO_{n-3}\to c_{1,\alpha_{1}}\CO^{22}_{(-1)^{n-2}\alpha_1}+\cdots~,
\end{eqnarray}
where the $c_{i,\alpha_j}$ are constants (by a rescaling we can set the coefficient of $\CO^{22}_{-\alpha_{n-2}}$ to one). The ellipses on the RHS of each of the \lq\lq$\to$" indicate possible (undetermined) mixings with various other dimension one monopole operators corresponding to other roots with the same charge under the RHS of \eqref{rmapA1A2m3} (and higher dimension monopole operators with the same charge under the $R$ symmetry, although such mixings presumably vanish by dimensional analysis). Note that, by our above arguments, there must at least be one non-trivial monopole operator of dimension one appearing on the RHS of each of the \lq\lq$\to$" in \eqref{OmapA1A2m3} (otherwise we would find that the operators whose vevs parameterize the Coulomb branch in four dimensions flow to operators whose vevs do not specify a subspace of the Coulomb branch in three dimensions). As a final comment, we observe that the $U(1)_R$ charges of the operators on the LHS of \eqref{OmapA1A2m3} agree with the charges of the operators on the RHS of \eqref{OmapA1A2m3} under the matching of symmetries in \eqref{rmapA1A2m3}.

Therefore, we see that the $\CN=2$ chiral generators of the four-dimensional theory are in one-to-one correspondence with the simple roots of the accidental $SU(n-1)$ Coulomb branch symmetry of the $S^1$ reduction. Moreover, the scaling dimensions of these generators are encoded in the quantum numbers of monopole multiplets that contain the conserved currents of the $SU(n-1)$ symmetry.\foot{Just as in the discussion below \eqref{OpmapA1A3}, the Weyl group acts on the $SU(n-1)$ quantum numbers of the operators on the RHS of \eqref{OmapA1A2m3} in the same way it acts on the roots.}

\newsec{$S^1$ reduction of the $(A_1, D_{2n})$ theory}
In this section, we will study the general $(A_1, D_{2n})$ theories. In the discussion around equation \eqref{IZS3A1D4}, we saw that, in the special case of the $(A_1, D_4)$ SCFT, the $S^3$ partition function of the $S^1$ reduction coincided with the $\beta\to0$ limit of the superconformal index (after dropping singular flavor-independent terms that measure a linear combination of the conformal anomalies of the four-dimensional theory \rcite{Buican:2015ina}). Furthermore, we argued that the reason for this matching was that the $(A_1, D_4)$ theory satisfies the quantization condition in \eqref{RQuantCond}.

However, for $n>2$, the $(A_1, D_{2n})$ SCFT violates \eqref{RQuantCond} since it has $n-1$ $\CN=2$ chiral generators of dimensions
\eqn{
E(\CO_{\ell})=2\left(1-{1\over 2n}\right)-{\ell\over n}~, \ \ \ 0\le\ell\le n-2~.
}[dimClopsD]
As a result, we expect that we will have non-trivial mixing of the $R$ symmetry and the topological symmetries in the three-dimensional limit.\foot{Note that the highest dimension generator has $E(\CO_0)=2\left(1-{1\over2n}\right)$. The remaining operators have precisely the same dimensions as the generators of the $(A_1, A_{2n-3})$ theory. Intuitively, this matching follows because, in the class $\CS$ construction, these latter generators are associated with the irregular singularity while the highest dimension generator is associated with the regular singularity \rcite{Xie:2012hs}. This intuition can be made more precise, because Higgsing the regular singularity induces a flow to the $(A_1, A_{2n-3})$ theory \rcite{Buican:2015ina}.} The $n>2$ case also differs from the $n=2$ case in one other important way: the flavor symmetry is $SU(2)\times U(1)$ instead of $SU(3)$.

In order to determine the expected mixing, we will first find it useful to rewrite the Schur index in \eqref{A1D2nSchur} as follows
\begin{eqnarray}\label{A1D2nindex}
\CI_{(A_1, D_{2n})}(q; a, b)&=&{\left(\CI_{\rm vect}^{SU(2)}(b)\right)^{-{1\over2}}\over(q;q)_{\infty}}\sum_{k=0}^{\infty}\Big({1\over b-b^{-1}}\sum_{s=\pm1}\left[{q^{n(2k+1)^2-2nk^2\over2}b^{2k+2}a^s\over1-q^{n(k+{1\over2})}ba^s}-{q^{n(2k+1)^2-2nk^2\over2}b^{-2k-2}a^s\over1-q^{n(k+{1\over2})}b^{-1}a^s}\right]\nonumber\\ &+&q^{nk(k+1)}\chi_{2k+1}(b)\Big)~,
\end{eqnarray}
Taking $\beta\to0$ with $q=e^{-\beta}$ and flavor fugacities $a=e^{-i\beta\zeta_1}$ and $b=e^{-i\beta\zeta_2}$, we find
\begin{eqnarray}\label{beta0limgenn}
\lim_{\beta\to0}\CI_{(A_1, D_{2n})}&\equiv& \CI^{4d\to3d}_{(A_1, D_{2n})}\simeq {1\over\sinh2\pi\zeta_2}\sum_{k=0}^{\infty}\sum_{s_1,s_2=\pm1}{(-1)^{s_2-1\over2}\over n(k+{1\over2})+i s_1\zeta_1+i s_2\zeta_2}\nonumber\\ &=&-{i\pi\over n}{1\over\sinh2\pi\zeta_2}\left(\tanh{\pi(\zeta_1+\zeta_2)\over n}-\tanh{\pi(\zeta_1-\zeta_2)\over n}\right)~.
\end{eqnarray}

\begin{table}
\begin{center}
     \begin{tabular}{| c | c | c | c | c | c |}
\hline   & $U(1)_1$ & $U(1)_2$ & $\cdots$ & $U(1)_{k-2}$& $U(1)_{n-1}$\cr\hline\hline
        $q_1$ & $1$ & $0$& $\cdots$ & 0 & 0\cr\hline
        $q_2$ & $1$ & $0$& $\cdots$ & 0 & 0\cr\hline
        $q_3$ & $1$ & $1$& $\cdots$ & 0 & 0\cr\hline
        $\vdots$ & $\vdots$ & $\vdots$& $\ddots$ & $\vdots$ & \vdots\cr\hline
        $q_n$ & $0$ & $0$& $\cdots$ & 1 & 1\cr\hline
        $q_{n+1}$ & $0$ & $0$& $\cdots$ & 0 & 1\cr\hline
      \end{tabular}
\caption{The matter fields (and charges) of the three-dimensional gauge theory that flows to the $S^1$ reduction of the $(A_1, D_{2n})$ theory.}
\label{AAsideD2n}
\end{center}
\end{table}
The $S^1$ reduction of the $(A_1, D_{2n})$ SCFT flows to an $\CN=4$ SCFT that is the IR limit of a $U(1)^{n-1}$ gauge theory with matter content summarized in Table \ref{AAsideD2n}. The theory also has hypermultiplet partners, $\tilde q^i$, with opposite quantum numbers under the $U(1)^{n-1}$ gauge symmetry.

Just as in the case of the $(A_1, A_{2n-3})$ theory, we will find it easier to work with the mirror theory \rcite{Xie:2012hs,Boalch:XXXXxx, Boalch:YYYYyy}. In this case, the mirror is the IR limit of a $U(1)^2$ gauge theory with matter content summarized in Table \ref{mirrorD2n}.
The $X_I$ with $I=1,\cdots, n-1$ have partners $Y^I$ of opposite gauge charges, and $A$ and $\hat A$ have partners $B$ and $\hat B$ with opposite gauge charges. Note that there is an $SU(n-1)$ flavor symmetry that acts on the $X_I$ via the fundamental representation and on the $Y^I$ via the anti-fundamental representation. Moreover, all the fields in Table \ref{mirrorD2n} are charged under a $U(1)$ flavor symmetry with charge $1/2$ (their partners have charge $-1/2$). These $SU(n-1)\times U(1)$ flavor symmetries are mapped to Coulomb branch symmetries in the direct $S^1$ reduction described by Table \ref{AAsideD2n}.

\begin{table}
\begin{center}
     \begin{tabular}{| c | c | c | c | c | c |}
\hline   & $U(1)_1$ & $U(1)_2$ \cr\hline\hline
        $X_I$ & $1$ & $1$\cr\hline
        $A$ & $1$ & $0$\cr\hline
        $\hat A$ & $0$ & $1$\cr\hline
      \end{tabular}
\caption{The matter fields in the mirror theory and their charges.}
\label{mirrorD2n}
\end{center}
\end{table}

The $S^3$ partition function for the mirror is
\eqn{
Z_{S^3}^{(A_1, D_{2n}),3dm}=\int d\sigma_1d\sigma_2{e^{\pi i(\xi_1\sigma_1+\xi_2\sigma_2)}\over\cosh\pi\sigma_1\cosh^{n-1}\pi(\sigma_1-\sigma_2)\cosh\pi\sigma_2}~,
}[ZS3A1D2n]
where we have suppressed the obvious dependence of the LHS on $\tilde u^i$ and $\tilde v^a$. We can again try to match the pole structures of \eqref{ZS3A1D2n} and \eqref{beta0limgenn} as we did in the case of $n=2$ using the identification
\eqn{
\zeta_1={1\over4}(\xi_1-\xi_2)~, \ \ \ \zeta_2={1\over4}(\xi_1+\xi_2)~.
}[D2nparid]
However, it is straightforward to check that this matching does not work for $n>2$.

As in the case of the $(A_1, A_{2n-3})$ theories, we can obtain $Z_{S^3}^{(A_1, D_{2n})}\simeq\CI_{(A_1, D_{2n})}^{4d\to3d}$ by turning on $SU(n-1)\to U(1)^{n-2}$ breaking imaginary partners of the real masses in the mirror. For even $n$ (with $n=2p$), we find
\begin{eqnarray}\label{ZS3A1D4n}
Z_{S^3}^{(A_1, D_{4p}),3dm}&=&\int d\sigma_1d\sigma_2{e^{\pi i(\xi_1\sigma_1+\xi_2\sigma_2)}\over\cosh\pi\sigma_1\cosh\pi\sigma_2\cosh\pi(\sigma_1-\sigma_2)}\nonumber\\&&\cdot{1\over\prod_{\alpha=1}^{p-1}\cosh\pi(\sigma_1-\sigma_2-{i\over2p}\alpha)\cosh\pi(\sigma_1-\sigma_2+{i\over2p}\alpha)}\\ &\simeq&-{i\pi\over 2p}{1\over\sinh\pi{\xi_1+\xi_2\over2}}\left(\tanh{\pi\xi_1\over 4p}+\tanh{\pi\xi_2\over 2p}\right)=\CI^{4d\to3d}_{(A_1, D_{4p})}~,\nonumber
\end{eqnarray}
while for odd $n$ (with $n=2p+1$), we have
\begin{eqnarray}\label{ZS3A1D4np2}
Z_{S^3}^{(A_1, D_{4p+2}),3dm}&=&\int d\sigma_1d\sigma_2{e^{\pi i(\xi_1\sigma_1+\xi_2\sigma_2)}\over\cosh\pi\sigma_1\cosh\pi\sigma_2}\nonumber\\&&\cdot{1\over\prod_{\alpha=1}^p\cosh\pi(\sigma_1-\sigma_2-i{2\alpha-1\over2(2p+1)})\cosh\pi(\sigma_1-\sigma_2+i{2\alpha-1\over2(2p+1)})}\\ &\simeq&-{i\pi\over 2p+1}{1\over\sinh\pi{\xi_1+\xi_2\over2}}\left(\tanh{\pi\xi_1\over 2(2p+1)}+\tanh{\pi\xi_2\over 2(2p+1)}\right)=\CI^{4d\to3d}_{(A_1, D_{2(2p+1)})}~.\nonumber
\end{eqnarray}

The above imaginary partners of the real masses describe (upon performing a mirror symmetry transformation) the following RG flow of the four-dimensional $U(1)_R$ charge
\eqn{
r\to I_3^{L,3d}+{1\over n\sqrt{2}}\sum_{a=1}^{n-2}(-1)^{n+a}{\sqrt{a(a+1)}}H_a~.
}[rmapA1D2m]
Note that the mixing in \eqref{rmapA1D2m} matches precisely the mixing in the $(A_1, A_{2n-3})$ theory described in \eqref{rmapA1A2m3}. In particular, the $H_a$ and corresponding weights are given as in \eqref{Cartan} and \eqref{weights} respectively (we again define the $n-2$ simple roots of $SU(n-1)$ to be $\alpha_a=\nu_a-\nu_{a+1}$). This matching is consistent with the fact that we can flow to the $(A_1, A_{2n-3})$ SCFT by Higgsing the regular singularity of the $(A_1, D_{2n})$ theory \rcite{Buican:2015ina} (note that this Higgsing preserves the Coulomb branch symmetries since we do not turn on vevs for $SU(2)_L$-charged operators).

Analogously to the case of the $(A_1, A_{2n-3})$ $S^1$ reduction, a particularly interesting set of operators parameterize the hyperk\"ahler Coulomb branch (in the flat space limit) via the following equations
\begin{eqnarray}\label{A1D2nOpEqs}
&&\CO^{22}_{0,-\alpha_a}\CO^{22}_{0,\alpha_a}=\CO^{22}_{0,0,a}\CO^{22}_{0,0,a+1}~,\nonumber\\ &&\CO^{22}_{0,\nu_{a_0}-\nu_{a_0+\ell}}\prod_{a=a_0+1}^{a_0+\ell-1}\CO^{22}_{0,0,a}=\prod_{a=a_0}^{a_0+\ell-1}\CO^{22}_{0,\alpha_a}~,\\ &&\CO^{222}_{-{1\over2},\nu_{n-1}}\CO^{22}_{0,\nu_i-\nu_j}=\CO^{222}_{-{1\over2},\nu_i}\CO^{22}_{0,\nu_{n-1}-\nu_j}\nonumber~, \\ &&\CO^{222}_{-{1\over2},\nu_{n-1}}\CO^{222}_{{1\over2}, -\nu_{i}}=\CO^{22}_{0,\nu_{n-1}-\nu_i}\left(\sum_j\CO^{22}_{0,0,j}\right)^2~,\nonumber
\end{eqnarray}
where the $\CO^{22}_{0,0,i}$ are the vector multiplet chiral scalars for the $U(1)^{n-1}$ gauge symmetry, the $\CO^{22}_{0,\alpha_a}$ are the monopole operators corresponding to the simple roots, the $\CO_{0,-\alpha_a}^{22}$ are the monopole operators corresponding to the reflections of the simple roots, the $\CO^{22}_{0,\nu_{a_0}-\nu_{a_0+\ell}}$ with $\ell>1$ are the monopole operators corresponding to the positive (non-simple) roots, and the $\CO^{222}_{\pm{1\over2},\pm\nu_i}$ are monopole operators of dimension $3/2$.\foot{There is an equation analogous to the second one in \eqref{A1D2nOpEqs} for the negative roots. Note also that under mirror symmetry, $\CO_{0,0,i}^{22}\leftrightarrow X_iY^i$, $\CO_{0,\alpha_a}^{22}\leftrightarrow X_aY^{a+1}$, $\CO_{0,-\alpha_a}^{22}\leftrightarrow X_{a+1}Y^{a}$, $\CO^{22}_{0,{\nu_{a_0}-\nu_{a_0+\ell}}}\leftrightarrow X_{a_0}Y^{a_0+\ell}$, $\CO^{222}_{{1\over2},-\nu_i}\leftrightarrow AY^i\hat A$, and $\CO^{222}_{-{1\over2},\nu_i}\leftrightarrow BX_i\hat B$. The operator relations discussed above can then be straightforwardly derived in the mirror theory (see also the recent discussion in \rcite{Bullimore:2015lsa}).} The first quantum number in the subscripts of the operators in \eqref{A1D2nOpEqs} is the charge under the $U(1)$ Coulomb branch flavor symmetry (this symmetry maps to a $U(1)$ flavor symmetry of the matter in the mirror discussed above \eqref{ZS3A1D2n}).

As in the case of the $(A_1, A_{2n-3})$ theory, we should demand that the $\CN=2$ chiral generators in four dimensions flow to operators in three dimensions whose vevs parameterize non-trivial subspaces of the Coulomb branch (in the flat space limit). Moreover, consistency of the operator maps with the flow to the $(A_1, A_{2n-3})$ theory requires that the $\CO_i$ with $i\ge1$ flow to linear combinations that include at least one monopole operator of dimension one. These requirements imply that
\begin{eqnarray}\label{OmapA1D2m}
&&\CO_1\to\CO^{22}_{-\alpha_{n-2}}+\cdots~, \ \CO_{2}\to c_{1,\alpha_{n-3}}\CO^{22}_{\alpha_{n-3}}+\cdots~, \cdots~, \CO_{k+1}\to c_{1,\alpha_{n-k-2}}\CO^{22}_{(-1)^{k+1}\alpha_{n-k-2}}+\cdots~,\nonumber\\&& \cdots~, \CO_{n-2}\to c_{1,\alpha_{1}}\CO^{22}_{(-1)^{n-2}\alpha_1}+\cdots~,
\end{eqnarray}
with the ellipses parameterizing mixing with other dimension one and higher monopole operators with the same $R$ charge (although mixings with higher dimensional operators presumably vanish on dimensional grounds). For the operator of the regular singularity, we find
\eqn{
\CO_0\to c_{+}\CO^{222}_{{1\over2},-\nu_{n-2}}+c_{-}\CO^{222}_{-{1\over2},\nu_{n-1}}+\cdots~, \ \ \ c_+\ne0 \ {\rm or} \ c_-\ne0~,
}[O0opmap]
where $c_{\pm}$ are undetermined constants, and the ellipses include possible mixing with higher dimension monopole operators (which again likely vanish).

As a result, we see that, just as in the case of the $(A_1, A_{2n-3})$ theory, the scaling dimensions of the $\CN=2$ chiral operators of the $(A_1, D_{2n})$ theory are encoded in the quantum numbers of the low-dimensional monopole operators in the $S^1$ reduction (in particular, for $n>3$, we see that the quantum numbers of the monopole operators associated with accidental symmetries in three dimensions play an important role).

\newsec{Comments on completing the operator map}
In \eqref{OpMapA1D4}, \eqref{OpmapA1A3final}, \eqref{OmapA1A2m3}, \eqref{OmapA1D2m}, and \eqref{O0opmap} we argued that certain linear combinations of monopole operators that partially parametrize the Coulomb branch in the three-dimensional IR SCFT (in the flat space limit) descend from $\CN=2$ chiral operators in the four-dimensional UV AD theory. Therefore, it is natural to ask if we can find a four-dimensional interpretation for the remaining operators that parameterize the Coulomb branch in three-dimensions (in the flat space limit). 

In general such a question is ill-defined. If we start from a well-defined operator at short distance, then it follows that this operator must flow to a well-defined operator at long distance. On the other hand, if we start with a well-defined operator in the IR, it need not come from a well-defined operator in the UV.

Still, we have seen that the four-dimensional $\CN=2$ chiral operators map to linear combinations of monopole operators that are related to the currents arising in the symmetry enhancement of the IR three-dimensional SCFT (or, in the case of the $\CO_0$ operator of the $(A_1, D_{2n})$ $S^1$ reduction, a monopole operator of the next-to-lowest dimension). As a result, it is tempting to imagine that the remaining linear combinations of monopole operators and IR descendants of three-dimensional vector multiplet scalars come from well-defined quantities in four dimensions that are part of some deeper structure of the parent AD theory.

While we do not have anything definite to say about this possibility, we can list the constraints on these potential ancestors (again, {\it assuming} they are well-defined, which need not be the case). A priori, the four-dimensional ancestors might be local or non-local.

If the four-dimensional ancestors are non-local, they could potentially be related to the line operators discussed in \rcite{Gaiotto:2010be}. If the four-dimensional operators are local, they may transform as parts of long multiplets or as parts of short multiplets. If they are part of long multiplets, we cannot say anything further. On the other hand, if the four-dimensional ancestors are part of local short multiplets (as in the case of the ancestors of the linear combinations described above), then we can say something about their possible superconformal representations.

Let us first consider the four-dimensional ancestors of the IR limits of the three-dimensional vector multiplet chiral scalars in the theories described in tables 1 and 2 (there are $n-2$ such operators in the $S^1$ reduction of the $(A_1, A_{2n-3})$ SCFT and $n-1$ such operators in the $S^1$ reduction of the $(A_1, D_{2n})$ SCFT). In three dimensions, these operators have $j_{L}=1$, $I_3^{L,3d}=-1$, and $j_{R}=0$. Furthermore, they are uncharged under the topological symmetries discussed above and so, by \eqref{SU2Lflow} and \eqref{SU2Rflow}, their four-dimensional ancestors should have $r=-1$ and $j_{R}=0$.

Clearly, there are no chiral operators with these quantum numbers in our AD theories, since they would correspond to free $U(1)$ multiplets. Moreover, such operators cannot be primaries in $\CB$, $\bar\CB$, or $\hat\CB$ type multiplets  since $R=0$ (these multiplets become chiral in this case).\foot{Here we are using the nomenclature of \rcite{Dolan:2002zh} (see also the earlier work \rcite{Dobrev:1985qv} which uses different terminology). The highest-weight primaries of the $\CB$ multiplets are annihilated by the $Q^1_{\alpha}$ supercharges (and $Q^2_{\alpha}$ as well if $R=0$; the $\bar\CB$ multiplets are the conjugate multiplets). The highest-weight primaries of the $\hat\CB$ multiplets are annihilated by $Q^1_{\alpha}$ and $\tilde Q_{2\dot\alpha}$. When $R=0$, the $\hat B$ mutltiplet is trivial since its heighest-weight primary is also annihilated by $Q^2_{\alpha}$ and $\tilde Q_{1\dot\alpha}$.} The $\hat\CC$ type multiplets are also ruled out since $r=-1$ implies that $j=1+\tilde j$ ($j$ and $\tilde j$ are the left-handed and right-handed Lorentz spins respectively) and so, upon reduction to three dimensions, such operators do not have scalar components.\foot{The spin zero $\hat\CC$ multiplets have highest weight components that are annihilated by $(Q^1)^2$ and $(\tilde Q_2)^2$ \rcite{Dolan:2002zh}. On the other hand, if the left-handed spin of the multiplet, $j$, is non-zero, then the spin $j-{1\over2}$ contraction with $Q^1_{\alpha}$ vanishes (and similarly for the right-handed spin, $\tilde j$, and the spin $\tilde j-{1\over2}$ contraction with $\tilde Q_{2\dot\alpha}$).} Our operators of interest in four-dimensions also cannot be primaries of scalar type $\CC$ multiplets since the highest weight primaries of these multiplets are annihilated by $(Q^i)^2$. In three dimensions, this property would imply that the adjoint chiral scalars are free, which is not correct (similar arguments rule out the higher-spin $\CC$ multiplets).

The only remaining possibility for a short local UV ancestor is a primary of type $\bar\CC_{0,-1(j_1,j_2)}$ (i.e., an operator that is annihilated by $(\tilde Q_i)^2$). These multiplets are captured by the index, but not by any of the special limits discussed in \rcite{Gadde:2011uv}. More generally, we can attempt to apply similar reasoning to the other $SU(2)_L$-charged operators that parameterize the Coulomb branch in the three-dimensional theory. However, we leave a more detailed investigation of such multiplets (and whether they actually have a sensible four-dimensional interpretation) to future work.

\newsec{Conclusions}
We have seen that the three-dimensional limits of the $(A_1, A_{2n-3})$ and $(A_1, D_{2n})$ theories contain some surprises. In particular, we saw that, when compactifying these theories on $S^1$, their $U(1)_R$ symmetries flowed to three-dimensional $R$ symmetries that mixed with topological symmetries of the corresponding $S^1$ reductions. Using this mixing we argued that the $\CN=2$ chiral primaries of the AD theories flowed to monopole operators of the three-dimensional descendants, and we saw new connections between the $U(1)_R$ quantum numbers of the four-dimensional $\CN=2$ chiral operators and the accidental symmetries of the reduction.

These results lead to some open questions. A partial list of these questions is as follows:

\begin{itemize}
    \item Find a method to compute the undetermined constants in \eqref{OpMapA1D4}, \eqref{OpmapA1A3final}, \eqref{OmapA1A2m3}, \eqref{OmapA1D2m}, and \eqref{O0opmap}.
    \item Understand if (some of) the remaining operators that parameterize the Coulomb branch of the $S^1$ reductions have well-defined four-dimensional interpretations. If so, are these local operators or non-local operators in four dimensions? Are they part of short multiplets or long multiplets?
    \item Is our $SU(2)$ quantization condition a biconditional statement? In particular, is it true that a theory satisfying the quantization condition must necessarily have no mixing of its $R$ symmetry with topological symmetries of the $S^1$ reduction? We saw some modest empirical evidence in favor of this statement in the $(A_1, D_4)$ example we studied.
    \item Generic AD theories (and, presumably, generic $\CN=2$ SCFTs) violate the $SU(2)$ quantization condition in \eqref{RQuantCond}. This fact implies that there should typically be non-trivial Coulomb branch symmetries in the corresponding $S^1$ reductions that mix with the $R$ symmetry. In the case of the $(A_1, A_{2n-3})$ and $(A_1, D_{2n})$ theories, these symmetries were topological $U(1)$ symmetries that enhanced to accidental non-Abelian symmetries in the IR. Does this enhancement always occur?    
    \item What does the existence of non-trivial Coulomb branch symmetries of the $S^1$ reduction of generic AD theories tell us about the space of $\CN=2$ SCFTs? Can we use properties of three-dimensional $\CN=4$ theories to say something general about the $\CN=2$ chiral spectra of $\CN=2$ SCFTs in four dimensions? Can we, perhaps, prove that $\CN=2$ SCFTs in four dimensions necessarily have rational dimensional $\CN=2$ chiral operators?
    \item On a related note, can we use any results found answering the questions in the previous item to show that $a$ and $c$ for $\CN=2$ SCFTs are necessarily rational? Since the rational numbers are countable, does the relation between four and three dimensions shed light on the nature of the resulting counting problem?
    \item Can we realize constraints similar to the ones discussed in this paper in flows between other dimensions?
\end{itemize} 

\ack{ \bigskip
We are grateful to D.~Kutasov, C.~Papageorgakis, L.~Rastelli, N.~Seiberg, and Y.~Tachikawa for interesting discussions and communications. We have also benefitted from discussions with  G.~Moore, C.~Papageorgakis, and D.~Shih on closely related topics. M.~B. would like to thank the members of the Queen Mary University of London CRST and the members of the University of Chicago EFI for stimulating scientific environments and discussions while parts of this work were being completed. Our research is partially supported by the U.S. Department of Energy under grants DOE-SC0010008, DOE-ARRA-SC0003883, and DOE-DE-SC0007897.
}

\newpage
\bibliography{chetdocbib}

\begin{thebibliography}{10}
\ifx\href\asklfhas\newcommand{\href}[2]{#2}\fi
\ifx\arxivref\asklfhas\newcommand{\arxivref}[2]{\href{http://arxiv.org/abs/#1}{#2}}\fi
\ifx\doiref\asklfhas\newcommand{\doiref}[2]{\href{http://dx.doi.org/#1}{#2}}\fi
\parskip 0pt
\normalsize

\bibitem{Argyres:1995jj}
P.C. Argyres \& M.R. Douglas,
\textit{``{New phenomena in SU(3) supersymmetric gauge theory}''},
\doiref{10.1016/0550-3213(95)00281-V}{Nucl.Phys. \textbf{B448}, 93 (1995)},
\normalsize{\texttt{\arxivref{hep-th/9505062}{hep-th/9505062}}}.

\bibitem{Argyres:1995xn}
P.C. Argyres, M.R. Plesser, N.~Seiberg \& E.~Witten,
\textit{``{New N=2 superconformal field theories in four-dimensions}''},
\doiref{10.1016/0550-3213(95)00671-0}{Nucl.Phys. \textbf{B461}, 71 (1996)},
\normalsize{\texttt{\arxivref{hep-th/9511154}{hep-th/9511154}}}.

\bibitem{Eguchi:1996vu}
T.~Eguchi, K.~Hori, K.~Ito \& S.K. Yang,
\textit{``{Study of N=2 superconformal field theories in four-dimensions}''},
\doiref{10.1016/0550-3213(96)00188-5}{Nucl.Phys. \textbf{B471}, 430 (1996)},
\normalsize{\texttt{\arxivref{hep-th/9603002}{hep-th/9603002}}}.

\bibitem{Gaiotto:2010jf}
D.~Gaiotto, N.~Seiberg \& Y.~Tachikawa,
\textit{``{Comments on scaling limits of 4d N=2 theories}''},
\doiref{10.1007/JHEP01(2011)078}{JHEP \textbf{1101}, 078 (2011)},
\normalsize{\texttt{\arxivref{1011.4568}{arXiv:1011.4568}}}.

\bibitem{Xie:2012hs}
D.~Xie,
\textit{``{General Argyres-Douglas Theory}''},
\doiref{10.1007/JHEP01(2013)100}{JHEP \textbf{1301}, 100 (2013)},
\normalsize{\texttt{\arxivref{1204.2270}{arXiv:1204.2270}}}.

\bibitem{Xie:2013jc}
D.~Xie \& P.~Zhao,
\textit{``{Central charges and RG flow of strongly-coupled N=2 theory}''},
\doiref{10.1007/JHEP03(2013)006}{JHEP \textbf{1303}, 006 (2013)},
\normalsize{\texttt{\arxivref{1301.0210}{arXiv:1301.0210}}}.

\bibitem{Buican:2014hfa}
M.~Buican, S.~Giacomelli, T.~Nishinaka \& C.~Papageorgakis,
\textit{``{Argyres-Douglas Theories and S-Duality}''},
\doiref{10.1007/JHEP02(2015)185}{JHEP \textbf{1502}, 185 (2015)},
\normalsize{\texttt{\arxivref{1411.6026}{arXiv:1411.6026}}}.

\bibitem{DelZotto:2015rca}
M.~Del~Zotto, C.~Vafa \& D.~Xie,
\textit{``{Geometric Engineering, Mirror Symmetry and 6d (1,0) $\to$ 4d,
  N=2}''},
\normalsize{\texttt{\arxivref{1504.08348}{arXiv:1504.08348}}}.

\bibitem{Buican:2014qla}
M.~Buican, T.~Nishinaka \& C.~Papageorgakis,
\textit{``{Constraints on chiral operators in $ \mathcal{N}=2 $ SCFTs}''},
\doiref{10.1007/JHEP12(2014)095}{JHEP \textbf{1412}, 095 (2014)},
\normalsize{\texttt{\arxivref{1407.2835}{arXiv:1407.2835}}}.

\bibitem{Gadde:2011ik}
A.~Gadde, L.~Rastelli, S.S. Razamat \& W.~Yan,
\textit{``{The 4d Superconformal Index from q-deformed 2d Yang-Mills}''},
\doiref{10.1103/PhysRevLett.106.241602}{Phys.Rev.Lett. \textbf{106}, 241602
  (2011)},
\normalsize{\texttt{\arxivref{1104.3850}{arXiv:1104.3850}}}.

\bibitem{Bonelli:2011aa}
G.~Bonelli, K.~Maruyoshi \& A.~Tanzini,
\textit{``{Wild Quiver Gauge Theories}''},
\doiref{10.1007/JHEP02(2012)031}{JHEP \textbf{1202}, 031 (2012)},
\normalsize{\texttt{\arxivref{1112.1691}{arXiv:1112.1691}}}.

\bibitem{Buican:2015ina}
M.~Buican \& T.~Nishinaka,
\textit{``{On the Superconformal Index of Argyres-Douglas Theories}''},
\normalsize{\texttt{\arxivref{1505.05884}{arXiv:1505.05884}}}.

\bibitem{Argyres:2012fu}
P.C. Argyres, K.~Maruyoshi \& Y.~Tachikawa,
\textit{``{Quantum Higgs branches of isolated N=2 superconformal field
  theories}''},
\doiref{10.1007/JHEP10(2012)054}{JHEP \textbf{1210}, 054 (2012)},
\normalsize{\texttt{\arxivref{1206.4700}{arXiv:1206.4700}}}.

\bibitem{Gadde:2011uv}
A.~Gadde, L.~Rastelli, S.S. Razamat \& W.~Yan,
\textit{``{Gauge Theories and Macdonald Polynomials}''},
\doiref{10.1007/s00220-012-1607-8}{Commun.Math.Phys. \textbf{319}, 147 (2013)},
\normalsize{\texttt{\arxivref{1110.3740}{arXiv:1110.3740}}}.

\bibitem{Dolan:2011rp}
F.~Dolan, V.~Spiridonov \& G.~Vartanov,
\textit{``{From 4d superconformal indices to 3d partition functions}''},
\doiref{10.1016/j.physletb.2011.09.007}{Phys.Lett. \textbf{B704}, 234 (2011)},
\normalsize{\texttt{\arxivref{1104.1787}{arXiv:1104.1787}}}.

\bibitem{Gadde:2011ia}
A.~Gadde \& W.~Yan,
\textit{``{Reducing the 4d Index to the $S^3$ Partition Function}''},
\doiref{10.1007/JHEP12(2012)003}{JHEP \textbf{1212}, 003 (2012)},
\normalsize{\texttt{\arxivref{1104.2592}{arXiv:1104.2592}}}.

\bibitem{Imamura:2011uw}
Y.~Imamura,
\textit{``{Relation between the 4d superconformal index and the $S^3$ partition
  function}''},
\doiref{10.1007/JHEP09(2011)133}{JHEP \textbf{1109}, 133 (2011)},
\normalsize{\texttt{\arxivref{1104.4482}{arXiv:1104.4482}}}.

\bibitem{Romelsberger:2005eg}
C.~Romelsberger,
\textit{``{Counting chiral primaries in N = 1, d=4 superconformal field
  theories}''},
\doiref{10.1016/j.nuclphysb.2006.03.037}{Nucl.Phys. \textbf{B747}, 329 (2006)},
\normalsize{\texttt{\arxivref{hep-th/0510060}{hep-th/0510060}}}.

\bibitem{Kinney:2005ej}
J.~Kinney, J.M. Maldacena, S.~Minwalla \& S.~Raju,
\textit{``{An Index for 4 dimensional super conformal theories}''},
\doiref{10.1007/s00220-007-0258-7}{Commun.Math.Phys. \textbf{275}, 209 (2007)},
\normalsize{\texttt{\arxivref{hep-th/0510251}{hep-th/0510251}}}.

\bibitem{Festuccia:2011ws}
G.~Festuccia \& N.~Seiberg,
\textit{``{Rigid Supersymmetric Theories in Curved Superspace}''},
\doiref{10.1007/JHEP06(2011)114}{JHEP \textbf{1106}, 114 (2011)},
\normalsize{\texttt{\arxivref{1105.0689}{arXiv:1105.0689}}}.

\bibitem{Aharony:2003sx}
O.~Aharony, J.~Marsano, S.~Minwalla, K.~Papadodimas \& M.~Van~Raamsdonk,
\textit{``{The Hagedorn - deconfinement phase transition in weakly coupled
  large N gauge theories}''},
\doiref{10.4310/ATMP.2004.v8.n4.a1}{Adv.Theor.Math.Phys. \textbf{8}, 603
  (2004)},
\normalsize{\texttt{\arxivref{hep-th/0310285}{hep-th/0310285}}}.

\bibitem{Jafferis:2010un}
D.L. Jafferis,
\textit{``{The Exact Superconformal R-Symmetry Extremizes Z}''},
\doiref{10.1007/JHEP05(2012)159}{JHEP \textbf{1205}, 159 (2012)},
\normalsize{\texttt{\arxivref{1012.3210}{arXiv:1012.3210}}}.

\bibitem{Hama:2010av}
N.~Hama, K.~Hosomichi \& S.~Lee,
\textit{``{Notes on SUSY Gauge Theories on Three-Sphere}''},
\doiref{10.1007/JHEP03(2011)127}{JHEP \textbf{1103}, 127 (2011)},
\normalsize{\texttt{\arxivref{1012.3512}{arXiv:1012.3512}}}.

\bibitem{Aharony:1997bx}
O.~Aharony, A.~Hanany, K.A. Intriligator, N.~Seiberg \& M.~Strassler,
\textit{``{Aspects of N=2 supersymmetric gauge theories in
  three-dimensions}''},
\doiref{10.1016/S0550-3213(97)00323-4}{Nucl.Phys. \textbf{B499}, 67 (1997)},
\normalsize{\texttt{\arxivref{hep-th/9703110}{hep-th/9703110}}}.

\bibitem{Kapustin:1999ha}
A.~Kapustin \& M.J. Strassler,
\textit{``{On mirror symmetry in three-dimensional Abelian gauge theories}''},
\doiref{10.1088/1126-6708/1999/04/021}{JHEP \textbf{9904}, 021 (1999)},
\normalsize{\texttt{\arxivref{hep-th/9902033}{hep-th/9902033}}}.

\bibitem{Buican:2013ica}
M.~Buican,
\textit{``{Minimal Distances Between SCFTs}''},
\doiref{10.1007/JHEP01(2014)155}{JHEP \textbf{1401}, 155 (2014)},
\normalsize{\texttt{\arxivref{1311.1276}{arXiv:1311.1276}}}.

\bibitem{Gaiotto:2009we}
D.~Gaiotto,
\textit{``{N=2 dualities}''},
\doiref{10.1007/JHEP08(2012)034}{JHEP \textbf{1208}, 034 (2012)},
\normalsize{\texttt{\arxivref{0904.2715}{arXiv:0904.2715}}}.

\bibitem{Gaiotto:2012xa}
D.~Gaiotto, L.~Rastelli \& S.S. Razamat,
\textit{``{Bootstrapping the superconformal index with surface defects}''},
\doiref{10.1007/JHEP01(2013)022}{JHEP \textbf{1301}, 022 (2013)},
\normalsize{\texttt{\arxivref{1207.3577}{arXiv:1207.3577}}}.

\bibitem{Nishioka:2011dq}
T.~Nishioka, Y.~Tachikawa \& M.~Yamazaki,
\textit{``{3d Partition Function as Overlap of Wavefunctions}''},
\doiref{10.1007/JHEP08(2011)003}{JHEP \textbf{1108}, 003 (2011)},
\normalsize{\texttt{\arxivref{1105.4390}{arXiv:1105.4390}}}.

\bibitem{Benini:2010uu}
F.~Benini, Y.~Tachikawa \& D.~Xie,
\textit{``{Mirrors of 3d Sicilian theories}''},
\doiref{10.1007/JHEP09(2010)063}{JHEP \textbf{1009}, 063 (2010)},
\normalsize{\texttt{\arxivref{1007.0992}{arXiv:1007.0992}}}.

\bibitem{Benvenuti:2011ga}
S.~Benvenuti \& S.~Pasquetti,
\textit{``{3D-partition functions on the sphere: exact evaluation and mirror
  symmetry}''},
\doiref{10.1007/JHEP05(2012)099}{JHEP \textbf{1205}, 099 (2012)},
\normalsize{\texttt{\arxivref{1105.2551}{arXiv:1105.2551}}}.

\bibitem{Beem:2013sza}
C.~Beem, M.~Lemos, P.~Liendo, W.~Peelaers, L.~Rastelli et~al.,
\textit{``{Infinite Chiral Symmetry in Four Dimensions}''},
\normalsize{\texttt{\arxivref{1312.5344}{arXiv:1312.5344}}}.

\bibitem{Boalch:XXXXxx}
P.~Boalch,
\textit{``{Irregular Connections and Kac-Moody Root Systems}''},
\normalsize{\texttt{\arxivref{0806.1050}{arXiv:0806.1050}}}.

\bibitem{Boalch:YYYYyy}
P.~Boalch,
\textit{``{Hyperkahler manifolds and nonabelian Hodge theory of (irregular)
  curves}''},
\normalsize{\texttt{\arxivref{1203.6607}{arXiv:1203.6607}}}.

\bibitem{Kapustin:2009kz}
A.~Kapustin, B.~Willett \& I.~Yaakov,
\textit{``{Exact Results for Wilson Loops in Superconformal Chern-Simons
  Theories with Matter}''},
\doiref{10.1007/JHEP03(2010)089}{JHEP \textbf{1003}, 089 (2010)},
\normalsize{\texttt{\arxivref{0909.4559}{arXiv:0909.4559}}}.

\bibitem{Bullimore:2015lsa}
M.~Bullimore, T.~Dimofte \& D.~Gaiotto,
\textit{``{The Coulomb Branch of 3d $\mathcal{N}=4$ Theories}''},
\normalsize{\texttt{\arxivref{1503.04817}{arXiv:1503.04817}}}.

\bibitem{Gaiotto:2010be}
D.~Gaiotto, G.W. Moore \& A.~Neitzke,
\textit{``{Framed BPS States}''},
\doiref{10.4310/ATMP.2013.v17.n2.a1}{Adv.Theor.Math.Phys. \textbf{17}, 241
  (2013)},
\normalsize{\texttt{\arxivref{1006.0146}{arXiv:1006.0146}}}.

\bibitem{Dolan:2002zh}
F.~Dolan \& H.~Osborn,
\textit{``{On short and semi-short representations for four-dimensional
  superconformal symmetry}''},
\doiref{10.1016/S0003-4916(03)00074-5}{Annals~Phys. \textbf{307}, 41 (2003)},
\normalsize{\texttt{\arxivref{hep-th/0209056}{hep-th/0209056}}}.

\bibitem{Dobrev:1985qv}
V.~Dobrev \& V.~Petkova,
\textit{``{All Positive Energy Unitary Irreducible Representations of Extended
  Conformal Supersymmetry}''},
\doiref{10.1016/0370-2693(85)91073-1}{Phys.Lett. \textbf{B162}, 127 (1985)}.

\end{thebibliography}

\end{document}